\documentclass[conference]{IEEEtran}
\IEEEoverridecommandlockouts

\usepackage{cite}
\usepackage{amsmath,amssymb,amsfonts}
\usepackage{algorithmic}
\usepackage{graphicx}
\usepackage{textcomp}
\usepackage{xcolor}
\usepackage{subcaption}
\usepackage{comment}
\def\BibTeX{{\rm B\kern-.05em{\sc i\kern-.025em b}\kern-.08em
    T\kern-.1667em\lower.7ex\hbox{E}\kern-.125emX}}
\begin{document}

\title{Prefetching in Deep Memory Hierarchies\\ with NVRAM as Main Memory}

\author{
     Manel Lurbe, Miguel Avargues, Salvador Petit, Maria E. Gómez, Rui Yang, Guanhao Wang and Julio Sahuquillo
     \IEEEcompsocitemizethanks{\IEEEcompsocthanksitem M. Lurbe, M. Avargues, S. Petit, M. E. Gómez and J. Sahuquillo are with the Departamento de Informática de Sistemas y Computadores, Universitat Polit\'ecnica de Val\'encia, Spain.
     E-mail: \{malursem, mavargues, spetit, megomez, jsahuqui\}@upv.edu.es
     \IEEEcompsocthanksitem Rui Yang and Guanhao Wang are with Huawei Technologies CO., LDT, China.
E-mail: \{yangrui50, guanhao.wang\}@huawei.com }%
}

\newif\ifremark
\long\def\remark#1{
\ifremark%
        \begingroup%
        \dimen0=\columnwidth
        \advance\dimen0 by -1in%
        \setbox0=\hbox{\parbox[b]{\dimen0}{\protect\em #1}}
        \dimen1=\ht0\advance\dimen1 by 2pt%
        \dimen2=\dp0\advance\dimen2 by 2pt%
        \vskip 0.25pt%
        \hbox to \columnwidth{%
                \vrule height\dimen1 width 3pt depth\dimen2%
                \hss\copy0\hss%
                \vrule height\dimen1 width 3pt depth\dimen2%
        }%
        \endgroup%
\fi}

\remarktrue

\newcommand{\julio}[1]{\remark{\color{blue}Julio: #1}}

\maketitle

\begin{abstract}
Emerging applications, such as big data analytics and machine learning, require increasingly large amounts of main memory, often exceeding the capacity of current commodity processors built on DRAM technology. To address this, recent research has focused on off-chip memory controllers that facilitate access to diverse memory media, each with unique density and latency characteristics. While these solutions improve memory system performance, they also exacerbate the already significant memory latency. As a result, multi-level prefetching techniques are essential to mitigate these extended latencies.

This paper investigates the advantages of prefetching across both sides of the memory system: the off-chip memory and the on-chip cache hierarchy. Our primary objective is to assess the impact of a multi-level prefetching engine on overall system performance. Additionally, we analyze the individual contribution of each prefetching level to system efficiency.
To achieve this, the study evaluates two key prefetching approaches: HMC (Hybrid Memory Controller) and HMC+L1, both of which employ prefetching mechanisms commonly used by processor vendors. The HMC approach integrates a prefetcher within the off-chip hybrid memory controller, while the HMC+L1 approach combines this with additional L1 on-chip prefetchers.

Experimental results on an out-of-order execution processor show that on-chip cache prefetchers are crucial for maximizing the benefits of off-chip prefetching, which in turn further enhances performance.
Specifically, the off-chip HMC prefetcher achieves coverage and accuracy rates 
exceeding 60\% and up to 80\%,
while the combined HMC+L1 approach boosts off-chip prefetcher coverage to as much as 92\%.
Consequently, overall performance increases from 9\% with the HMC approach to 12\% when L1 prefetching is also employed.
\end{abstract}


\section{Introduction}
\label{sec:introduction}
A large set of emergent applications, like big data or machine learning methods, require huge amounts of main memory which exceeds the memory capacity of existing commodity processors \cite{Mak18}. To cope with the increasing memory needs, two main design alternatives have been followed. 

The typical design choice focuses on the improvement of the DRAM memory subsystem in terms of storage capacity and memory bandwidth with each new generation. DRAM has scaled (in Mbits$/$chip) in a $2\times$ factor every 1.5 years from 1985. 
However, this trend has slowed down since the beginning of this century and becomes challenging below the 10nm technology node since the implementation becomes unreliable \cite{Meza15, Mutlu16}. 
To address this shortcoming and the growing memory bandwidth demands, current server processors have increased the number of memory channels attached to the memory controller to support higher memory-level parallelism (MLP).
For instance, the Intel Xeon Platinum 8470N Processor \cite{IntelXeonPlatinum8470} launched in Q1'23 is equipped with eight memory channels.
This design choice, however, does not scale due to the pin count constraint of current chips since a huge number of pins are required to connect the processor to the increasing number of channels. 

The second design choice concentrates on the use of alternative denser and more energy-efficient RAM technologies \cite{Burr08}. In this regard, the use of NVRAM (non-volatile RAM) has been explored both in the academia and the industry \cite{Optane19, Optane21}. Many papers have focused on this technology, both working as main memory and as a fast secondary memory. NVRAM can provide more than 500GB in a single DIMM (called NVDIMM). This means that if NVRAM is implemented as main memory \cite{Hoya19}, just a few NVDIMMs would provide enough capacity for applications that demand a large amount of memory space.


A significant drawback of the latter approach, as Intel has acknowledged \cite{Inteloptanebusiness}, is that the manufacturing technology remains immature. This makes it impractical for Intel to produce and deliver products at the required scale, especially as a single-source supplier.
In fact, Intel cited this as the primary reason for discontinuing the Intel Optane product line \cite{Optane19, Optane21}. 
Of course, company earnings fuel product development;
however, one might expect that advances in manufacturing technology will make NVRAM products affordable enough to be cost-effective in the near future.
In alignment with Intel, which acknowledges the exceptional potential of this technology \cite{Inteloptanebusiness}, we believe it is valuable to explore NVRAM as main memory from an academic perspective.

From an architectural perspective, the main disadvantage of using NVRAM as main memory is its longer access time compared to DRAM technologies \cite{Pel14,Zuo18,Luo18,Kagar22}, which would translate into poor and unacceptable applications' performance. 
Because of NVRAM latencies, using this technology as main memory cannot be done straightforwardly. 
Instead, its use introduces important performance implications, so that the memory hierarchy should be rethought and re-architected. In other words, the memory structures of the memory hierarchy must be first defined to provide a \emph{collaborative} cache hierarchy capable of providing a huge storage capacity while hiding the huge memory access times. 
In this context, distinct memory organizations are possible \cite{Avargues23}. To focus the research, this paper assumes a hybrid memory hierarchy,
composed of SRAM, DRAM and NVRAM devices,
closely resembling that deployed in the Intel Optane \cite{Optane19, Optane21} (see Section \ref{sec:organization} for further details).

Several previously published works propose distinct hybrid memory approaches to hide the long NVRAM access times \cite{rowbuffermigration,pdram,qureshi2009scalable,loh11,DRAMTags,transformer,hybrid2}.
However, the effect of prefetching is commonly omitted despite its significant impact on the overall performance. This paper analyzes the impact of prefetching across different levels of the hierarchy in such memory systems. To the best of our knowledge, this issue has not been explored in prior research.

Experimental results, for a state-of-the-art out-of-order (o-o-o) execution processor, show that prefetching is able to significantly hide NVRAM latencies at the MC cache. However, minor performance benefits can be appreciated in the overall performance. This observation led us to implement an additional prefetching engine, located at the L1 processor cache, to close the gap between the latencies of the L1 processor cache and the MC cache. Results show that the impact of this two-level prefetching system can improve L1 miss latencies between 40\% and 50\%, which translates into overall performance improvements 
over 8\% and up to 12\%.

This paper makes three main contributions: 
i) we present a detailed analysis of prefetching impacts across the whole (on-chip and off-chip) hybrid memory hierarchy;
ii)	we found that the prefetching performance at the HMC cache is really impressive in terms of both accuracy and coverage; iii) we claim that multiple prefetchers along the cache hierarchy are needed to translate the achieved HMC cache prefetching benefits into overall system performance gains in a state-of-the-art out-of-order processor.

The remainder of this paper is organized as follows.
Section \ref{sec:background} presents some background on commercial prefetchers and the baseline hybrid memory organizations.
Section \ref{sec:prefetcher-design-choices} summarizes the memory controller and its implications on prefetching.
Section \ref{sec:framework} discusses the simulation framework.
Section \ref{sec:performance evaluation} presents and analyzes the results.
Section \ref{sec:related} discusses related work.
Finally, Section \ref{sec:conclusions} draws some concluding remarks.

\section{Background}
\label{sec:background}

This paper analyzes the impact of common prefetchers deployed in commercial processors on hybrid (DRAM + NVRAM) memory systems.
To provide context to understand the analysis, this section presents a summary of relevant background concepts, first on prefetching in commercial processors and then on the hybrid memory organization used as baseline in this work.

\subsection{Prefetchers in Commercial Processors}
\label{sec:prefetchers}

Although many prefetching techniques have been proposed in academic research \cite{bhatia19, bakhshalipour19, shevgoor15, pakalapati20, michaud16, kim16, navarro22}, often involving complex hardware structures, in practice, most commercial processors rely on two main types of prefetchers due to their low complexity and cost-effectiveness: \emph{next-line} and \emph{stride} prefetchers. 
Since our focus is on commercial implementations, 
this section summarizes some next-line and stride prefetcher information published by processor manufacturers.

\subsubsection{Next-Line Prefetchers}

Next-line prefetchers, also known as sequential, adjacent, or one-block-lookahead (OBL), prefetch adjacent cache lines.
Next-line prefetching is the simplest form of prefetching and is commonly implemented along the cache hierarchy of modern processors.
The main aim of this prefetcher is to exploit spatial locality, which is especially high when the processor accesses to instructions of the program code. In addition, many applications make use of regular data structures (e.g., large data arrays) so that they exhibit a rich spatial locality. Therefore, this kind of prefetcher is implemented to prefetch both instructions and data.

There are different variants of next-line prefetching, which mainly differ on when the prefetch is triggered: i) only on cache misses, ii) always, and iii) selectively (e.g. on prefetch hits).
When prefetching instructions, this prefetcher only considers an ascending order of memory addresses; in contrast, in data caches, it can follow both ascending or descending order.

In Intel and ARM processors, next-line prefetchers can be found alongside different cache levels (e.g. in the L1 and the L2 caches). For instance, in the 
ARM Cortex-A57 MPCore, L1 I-cache next-line prefetching works as follows: on a cache miss, the next line is searched for in the same cache; if the next line is also missing, a subsequent L2 lookup is initiated \cite{arm-L1Icache-next-line}.
Regarding the L2 cache, the Cortex-A57 incorporates a mechanism to identify and prefetch instruction streams. This mechanism retrieves consecutive cache lines when there is an L2 instruction fetch miss or an L2 cache prefetch hit \cite{arm-L2-next-line}. 

Intel typically does not disclose the details of its hardware prefetch mechanisms, but there are some exceptions. 
For example, in \cite{corporation2018intel} some details are provided. 
It is stated that the L1 streaming prefetcher (also known as the \emph{DCU prefetcher}) fetches one line beyond an ascending stream of addresses. In addition, the \emph{L2 streaming prefetcher} can issue either one or two prefetches in response to each L2 cache lookup.
Detected streams are prefetched either in increasing or decreasing address order.

\subsubsection{Stride Prefetchers}

Stride prefetchers detect access patterns where sequences of memory addresses follow a constant difference, or stride.
This type of prefetching is highly accurate when, for example, the elements of a large, regular structure, such as a multidimensional array, are accessed.

Typical stride prefetching implementations employ a simple hardware table in the L1 data cache to detect the stride. For each entry in the table, three main fields are implemented \cite{tabla-stride}: i) the IP (or a tag) to identify the load instruction to be tracked in the entry, ii) the address of the last access performed by the load, and iii) the stride, calculated as the difference between two sequential addresses issued by the load.
Upon a new access of a load instruction, the control logic calculates the difference between the address of the current access and the last one stored in the corresponding entry of the table. If the computed difference matches the stored stride, a stride hit is detected. Thus, a prefetch is triggered, and a stream buffer is allocated to prefetch the stream.

As next-line prefetchers, stride prefetchers are also implemented at different cache levels of commercial processors.
For instance, in the Cortex-A57, \emph{load-store} streams are detected in the L2 cache \cite{arm-L2-next-line}.
Concerning Intel, according to \cite{corporation2022intel},
the \emph{Instruction Pointer Prefetcher} (IPP) keeps track of load instructions and generates prefetching requests in the L1 cache if a stride (forward or backward) access pattern is detected.

\subsection{Hybrid Memory System Organization}
\label{sec:organization}

\begin{figure}[t] 
\begin{center}
\includegraphics[width=0.50\columnwidth]{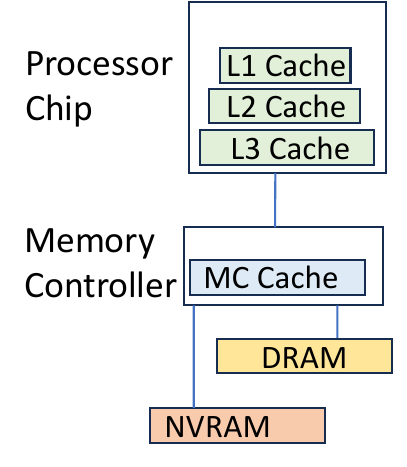}
\caption{Hybrid memory system organization.}
\label{fig:memory-hierarchy}
\end{center}
\end{figure}

The main aim of hybrid memory organizations is to architect the memory hierarchy with distinct types of memory media to provide a huge storage capacity while hiding the involved long NVRAM latencies.
Distinct organizations are possible by adding conventional technologies (i.e., SRAM and DRAM) in the path to the NVRAM-based main memory to achieve the best tradeoff between storage and speed.

An interesting organization devised by Intel \cite{Optane19, Optane21} is to add a DRAM structure as a cache of the NVRAM. 
This way allows 
much lower latency than directly accessing the NVRAM main memory.
Nevertheless, employing DRAM as a cache brings its own set of challenges. Firstly, the memory controller must efficiently manage data transfers between the NVRAM main memory and the DRAM cache. Secondly, DRAM access is destructive, as capacitor charges are lost with each access, which complicates the management of DRAM media as a cache. Additionally, DRAM access latency can still be considered unacceptable for high-performance processors. 

To deal with these shortcomings, an additional SRAM cache level is introduced in \cite{Avargues23}. This cache is designed to be large enough to hide the long latencies of DRAM and NVRAM accesses, while also reducing the complexity of managing various memory media.
Figure \ref{fig:memory-hierarchy} presents this hierarchy, which consists of two main parts from the processor perspective: on-chip and off-chip. The former includes the memory structures deployed in the processor chip referred to as on-chip hierarchy, which connects with the latter through a memory controller (MC).
The MC includes an SRAM cache (MC cache), and controls hybrid memory media, composed of DRAM and NVRAM devices. The NVRAM acts as the main memory, and the DRAM operates as a cache of the NVRAM. 

From now on, this organization, referred to as NV-S-D (which stands for NVRAM plus SRAM cache plus DRAM cache), will act as the baseline for this paper.

\section{HMC Design and Implications on Prefetching
}
\label{sec:prefetcher-design-choices}

This section summarizes the main features of the hybrid memory controller to help understand the prefetchers' efficiency and discusses the prefetchers location alongside the cache hierarchy.
 Finally, we summarize the deployed prefetching systems, the attached caches, and the metrics used to evaluate the prefetching performance.

\subsection{Hybrid Memory Controller}

As explained above, the NV-S-D organization introduces a two-level off-chip cache hierarchy.
In this organization, the memory controller is a critical system component, as it manages two caches with distinct technologies and the NVRAM media. Since it handles distinct memory technologies, we refer to it as the \emph{Hybrid Memory Controller} or HMC.
The memory controller was originally proposed in \cite{Avargues23}, but it is summarized in this section to make the paper self-contained.

The SRAM cache in the HMC and the DRAM cache in commodity DIMMs are the two caches handled by the HMC.
The former is implemented as a sector cache called the \emph{HMC cache}.
Sector caches allow storing multiple data blocks in a single large cache line (e.g., a 256B sector cache line allows storing four 64B data blocks). This way reduces the number of accesses to the lower levels of the HMC hierarchy.
Its size is relatively small (e.g., 64 MB) compared to the DRAM cache capacity.
In other words, by design, the HMC sector cache acts as a sequential prefetcher that helps hide the latency of the slow NVRAM media. 

A key question in the design of the DRAM cache is how tags are handled. 
Since DRAM reads are destructive, we implement tags in an auxiliary fast SRAM table in the HMC.
This table is referred to as the \emph{tag cache} as it keeps tags of the blocks stored in the DRAM cache.
In other words, the tag cache keeps the tag array of the data blocks stored in the DRAM cache (which stores only the data array). This design accesses a data block in the DRAM cache only in case of a tag hit in the tag cache. Notice that accessing the data array of the DRAM cache upon every sector cache miss would result in significant refreshing overhead due to a large number of DRAM cache misses.
To deal with this shortcoming, the much faster tag cache is accessed first to check whether the requested sector is available in the data array (DRAM), and only on a tag hit is the DRAM accessed.
See \cite{Avargues23} for further details.

\subsection{Implemented Prefetching Systems and Locations}
\label{sec:pref-systems-locations}

Two main prefetching systems have been studied and implemented in two main cache structures along the cache hierarchy. The former (off-chip) system implements the prefetching mechanisms in the HMC cache and is referred to as the HMC prefetching system or HMC system. The latter (on-chip \& off-chip) complements the former by also prefetching in the L1 data cache and is known as the HMC + L1 (prefetching) system. Below, we discuss these prefetching systems and the reasons why we devised them.

\subsubsection{HMC System}

As the primary goal of the prefetching systems devised in this work is to hide the long latency of accessing the NVRAM main memory in the NV-S-D system, 
we placed the first system as close as possible to this memory to check its effectiveness.

The memory structure closest to the main NVRAM memory is the DRAM cache, which works as the second-level off-chip cache.
However, making this cache store prefetches would introduce significant refreshing overhead. 
Notice also that these prefetches would incur the replacement of previous DRAM cache blocks, which, due 
to the working nature of DRAM (e.g., limited bandwidth and refreshing), would worsen the performance even more. 
For these reasons, we opted to store the prefetches in the HMC cache.

This system works as follows: when a prefetch is triggered by the HMC cache prefetching logic,
the tag cache is looked up to check if the requested block is stored in the DRAM data array.
On a tag hit, the DRAM is accessed; otherwise, the requested block is brought from the NVRAM cache 
and copied to the HMC cache.

We 
study first the impact of implementing a prefetching system in the HMC cache on the NVRAM latency. This is expected to help reduce the HMC cache miss rate and improve overall memory latency by bringing blocks in advance from the NVRAM to the HMC cache.

\subsubsection{HMC+L1 System}

This prefetching system extends the previous one by adding prefetching logic to the L1 processor cache.
The HMC system can reduce the NVRAM latency by increasing the number of HMC cache hits due to the blocks fetched in advance. Thus, it is a good starting point. 
However, as experimental results will show, achieving this milestone does not necessarily translate into significant gains in the overall system performance for all the workloads.
The key reason is that although the HMC system significantly reduces off-chip memory latency, it is still above one hundred processor cycles, which introduces serious constraints on the overall system performance achieved by aggressive out-of-order processors. Consequently, this scenario still limits the potential performance gains introduced by prefetching mechanisms.


To address this situation, the average memory latency (as seen from L1) should be improved. 
To this end, prefetching mechanisms can be deployed along the whole cache hierarchy (e.g., the L1 and the L2 caches). To prove this claim, we implemented, in addition to the HMC cache prefetcher, a prefetching engine in the L1 data cache. Of course, performance could be further enhanced by adding prefetchers at intermediate on-chip cache levels of the cache hierarchy (e.g., L2); however, this refinement is out of the scope of this paper.

\subsubsection{Studied Prefetchers}

We used prefetchers commonly implemented in current processors (next-line and stride, see Section \ref{sec:prefetchers}) for the prefetching engines studied in this work.
Both prefetchers are included in the L1 cache, 
whereas the HMC cache only implements the next-line prefetcher, since the stride prefetcher showed negligible impact on the performance of this cache.
We looked into the reasons and found that the stride prefetcher rarely detects streams at the HMC, mainly due to stride access patterns being filtered out by hits in the upper cache levels. 

\subsection{Prefetching Evaluation Metrics}
\label{prefetcher-evaluation-metrics}

Prefetcher performance is generally measured using two key metrics: coverage and accuracy. Coverage indicates the percentage of cache misses avoided due to prefetching, relative to the total number of misses without prefetching. Accuracy, on the other hand, measures the percentage of prefetched blocks that are actually requested by the processor.
A good prefetcher should provide both good coverage and accuracy in order to improve the overall system performance.

Accuracy and coverage have been computed using Equation \ref{eq. accuracy} and Equation \ref{eq. coverage}, respectively;
where
$Misses_{with\_Pref}$ and $Misses_{without\_Pref}$
refer to demand cache misses experienced in a system with prefetch and with no prefetch, respectively; and 
$\# of Issued\_Prefetches$ refers to the number of triggered prefetches.

\begin{equation}
    Accuracy = \frac{Misses_{without\_Pref} - Misses_{with\_Pref}}{\#Issued\_Prefetches} \label{eq. accuracy}
\end{equation}

\begin{equation}
    Coverage = 1 - \frac{Misses_{with\_Pref}}{Misses_{without\_Pref}} \label{eq. coverage}
\end{equation}


\section{Simulation Framework}
\label{sec:framework}


\subsection{Simulation Tools}
\label{subsec:simulator}

Experiments have been carried out with the gem5 event-driven simulator and the NVMain simulator. 
Gem5 has been used to simulate the processor and the on-chip cache hierarchy. Experiments have been run with the 
full system simulation mode, where the simulated system behaves much like a real machine, starting from boot.
NVMain is a cycle-accurate main memory simulator able to model both NVRAM and DRAM memories of the off-chip cache subsystem, including the hybrid memory and the memory controller.

\subsection{Workloads}
\label{subsec:workload}


Three main workloads, Redis, Memtier, and MySQL, commonly used in big data environments for database management and data processing, have been analyzed in this work. These applications are widely adopted in modern data-centric architectures due to their relevance in handling large-scale data operations. The main characteristics of these applications are described below.

\textit{Redis} is an in-memory storage structure, which can be used as a database, cache, or message broker. 
It is known for its high performance and low latency, which are crucial characteristics in big data applications to accelerate frequent queries or handle large volumes of real-time data, and has been used in recent works \cite{bajaber20}.
It uses a wide variety of data structures such as hashes, lists, sets, ordered sets, etc. This workload natively supports data replication, the LRU (least recently used) replacement policy, and multiple on-disk persistence levels. It is composed of two main benchmarks, discussed next: 

  
\textit{Redis-benchmark} is a utility program included in the redis installation that offers a way to simulate commands as if they were committed by N clients at the same time for a total of M requests.

\textit{Memtier-benchmark} 
is a tool commonly used to benchmark the performance of in-memory storage systems like Redis or Memcached. It allows simulation of realistic workloads typical of big data environments.
Some of its features are a configurable read / write ratio and the ability to choose the data access pattern.


\textit{MySQL} is one of the most well-known relational database management systems (DBMS) used in various applications that handle large volumes of data to store and query structured data \cite{panda17}. For instance, it is used by giant technology companies such as Google, Twitter, and Facebook \cite{dawodi19}. Its code is open source and provides a wide variety of key features such as rollback and fault recovery. 
We use the sysbench benchmarking tool to perform the experiments with the MySQL server.

To choose the configuration parameters of each workload to carry out the evaluation study, we first launched a wide set of experiments to 
check that the parameters' values were properly stressing the memory hierarchy.
Figure \ref{fig:benchmark-params} summarizes the parameter values for each workload.

\subsection{System Parameters}
\label{subsec:system}

This section presents the main system parameters for the NV-S-D system model split into three major subsystems: the memory hierarchy, the processor cores, and the prefetcher engine.

\subsubsection{Memory Hierarchy}

\begin{figure}[t]
     \centering
     \subfloat[Redis]{
     \scriptsize
         \raisebox{0.7cm}{
             \begin{tabular}{|c|c|}
             \hline
             \textbf{redis' } & \\
             \textbf{Parameter} & \textbf{Value} \\\hline
             Clients     & 200 \\\hline
             Requests    & 45000 \\\hline
             Size        & 2000 B \\\hline
             Tests       & get, set, lpush, lpop \\\hline
             \end{tabular}
         }
     }\hspace{0.2cm}
     \subfloat[Memtier]{
     \scriptsize
         \raisebox{0.2cm}{
             \begin{tabular}{|c|c|}
             \hline
             \textbf{memtier's } & \\
             \textbf{Parameter} & \textbf{Value} \\\hline
             Threads     & 4 \\\hline
             Clients     & 15 \\\hline
             Data size   & 1024 \\\hline
             Ratio       & 0:1 \\\hline
             Key min/max & 1/200000\\\hline
             Key pattern & R:R\\\hline
             Pipeline & 32 \\\hline
             Test time & 60 s\\\hline
             \end{tabular}
         }
     }\\\vspace{0.1cm}
     \subfloat[MySQL]{
     \scriptsize
         \raisebox{0.8cm}{
             \begin{tabular}{|c|c|}
             \hline
             \textbf{MySQL's } & \\
             \textbf{Parameter} & \textbf{Value} \\\hline
             Threads & 8\\\hline
             Table size & 100000\\\hline
             Db driver     &  mysql\\\hline
             Time   & 30s \\\hline
             Max requests & 0 \\\hline
         \end{tabular}
         }
     }
     \caption{Parameters used for each workload.}
     \label{fig:benchmark-params}
\end{figure}

In NV-S-D system, the off-chip memory includes three major components: the HMC sector cache, the DRAM cache, and the NVRAM media. The on-chip cache hierarchy comprises the L1 and the L2 processor caches.
Table \ref{tab:gem5_system_config} summarizes each component's main parameters, which include the geometry (e.g., cache size, block size, and number of cache ways) and access latency. These values were obtained for a 22nm technology node with Hewlett Packard's CACTI 7.0 \cite{cacti7.0} memory and cache modeling tool.

\subsubsection{Modeled Out-of-Order Processor and Prefetcher Parameters}

\begin{table}[t]
\scriptsize
\centering
\caption{Gem5 and NVMain system configuration.} 
\label{tab:gem5_system_config}
\resizebox{0.95\columnwidth}{!}{  
\begin{tabular}{|c|l|c|}
\hline
Component & Parameter & Value\\\hline\hline
L1 data cache & Size, block size, associativity  & 32 KB, 64 B, 8-way \\\cline{2-3}
              & Tag/Data latency & 3/3 cycles
    \\\hline
L1 inst. cache & Size, block size, associativity  & 32 KB, 64 B, 8-way \\\cline{2-3}
               & Tag/Data latency & 3/3 cycles 
      \\\hline
L2 cache      & Size, block size, Associativity  & 2 MB, 64 B, 16-way \\\cline{2-3}
              & Tag/Data latency & 11/11 cycles\\\cline{2-3}
              & Write policy & Write-back
       \\\hline\hline
HMC sector cache & Size, block size, associativity  & 8 MB, 256 B, 16-way \\\cline{2-3}
                & Tag/Data latency & 17/17 cycles\\\cline{2-3}
                & Write policy & Write-back
        \\\hline\hline
DRAM cache      & Memory channels  & 2  \\\cline{2-3}
                & Total space storage  & 64MB \\\hline\hline
DRAM Media  & Read/Write Access Time  & 33/11 Cycles \\\hline\hline
NVRAM Media & Read/Write Access Time  & 353/86 Cycles \\\hline
\end{tabular}
}
\end{table}

\begin{table}[t]
\scriptsize
\centering
\caption{Main parameters for the DerivO3CPU.}
\label{tab:o3-params}
\resizebox{0.65\columnwidth}{!}{  
\begin{tabular}{|l|r|}\hline
Parameter & Value \\\hline\hline
ROB size (\# of Entries) &  192\\ \hline
IQ size (\# of Entries) &   64\\\hline
Integer Registers &  256\\ \hline
Floating-Point Registers &  256\\\hline\hline
Integer ALU Units &  6\\\hline
Integer Divider \& Multier Units &  2\\\hline
Floating-Point ALU Units &  4\\\hline
Floating-Point Divider \& Multiplier Units &  2\\\hline\hline
SIMD Units & 4\\\hline\hline
Functional Units (Operators) Latency & 1\\\hline
\end{tabular}
}
\end{table}


Experiments were launched with an aggressive out-of-order (o-o-o) processor, which allows for the introduction of higher memory-level parallelism. Notice that an in-order processor would not be appropriate for this study since it implements locking caches, which sequences the access to the main memory.
Because of this reason, and to provide representative results, we used an o-o-o processor ("DerivO3CPU" gem5 model) configured with an aggressive superscalar 6-instruction issue width.
Table \ref{tab:o3-params} presents key microarchitectural parameters for the o-o-o processor closely resembling state-of-the-art processors. For simplification purposes, the latency of all the functional units is assumed to be one cycle, as typically considered in the literature. 

Regarding the next-line and stride prefetchers, they are configured with a depth of 2 and 4 respectively, and both 
implement 8 32-entry stream buffers.
As mentioned in Section \ref{sec:pref-systems-locations}, these prefetchers are included in the L1 cache while the HMC cache only uses the next-line prefetcher.

\section{Performance Evaluation}
\label{sec:performance evaluation}

This section evaluates the impact on performance of the proposed prefetching systems.
The evaluation is made in a refined way, studying step by step, how the performance at different levels of the system is affected.
First, we evaluate the performance of the system themselves with specific metrics such as accuracy and coverage (see Section \ref{prefetcher-evaluation-metrics}).
After that, we analyze to what extent improving these metrics enables reducing the number of cache misses in each cache of the hierarchy. That is, how the prefetchers are capable of improving the cache hierarchy performance, making special focus on the experienced cache latency.
Finally, we evaluate the impact of the studied prefetching approaches on the overall system performance.

\begin{figure}[t]
    \centering
    \includegraphics[width=0.3\columnwidth,trim={25cm 33.8cm 25cm 1cm}, clip]{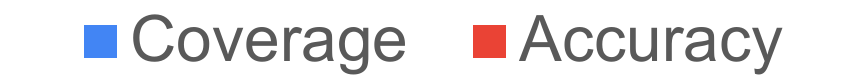}
    \includegraphics[width=\columnwidth,trim={0 0 0 3cm},clip]{figures/o3_results/cov_acc_o3.pdf}
    \caption{Performance (coverage and accuracy) of the implemented prefetchers  across the studied benchmarks.}
    \label{fig:acc-cov-pf-stats}
\end{figure}

\subsection{Prefetching Performance Evaluation} 

This section evaluates 
the accuracy and coverage of 
the devised HMC and HMC+L1 prefetching systems. Figure \ref{fig:acc-cov-pf-stats} presents the results for 
the implemented systems according to the cache where prefetchers are allocated.
That is, attached to the HMC cache in the HMC system, and both in the HMC and L1 caches in the HMC+L1 system. Results are provided for the three studied workloads (\texttt{redis}, \texttt{memtier}, and \texttt{MySQL}). Three important observations can be made.

First, it can be appreciated that the HMC system achieves significant performance (both coverage and accuracy) gains across the three studied applications, especially in \texttt{memtier} where both metrics are over 73\%. This means that the HMC system effectively reduces the amount of HMC cache misses with little  bandwidth increase (as it achieves both high accuracy and coverage).

Second, in the HMC+L1 system, the prefetching at the HMC cache presents interesting performance differences with respect to the observed in the HMC system. More precisely, it can be appreciated that coverage improves while accuracy slightly drops. The main reason is that HMC cache in the HMC+L1 system receives more requests than that of the HMC system because the latter system generates additional prefetches from the L1 processor cache.

Third, the HMC+L1 system offers lower coverage and accuracy at the L1 cache, but even so, both metrics are over 30\% with the only exception of coverage for \texttt{MySQL}, which is around 26\%.  These figures, as latency results will show, will help improve overall performance.

In summary, adding a prefetching engine to the L1 cache (i.e. HMC+L1 prefetching system) improves coverage with respect to the HMC system despite there being a slight drop in accuracy.

\subsection{Memory Hierarchy Performance Analysis} 
\label{cap:mem_hierarchy_perf}

\begin{figure*}[t]
    \centering
    \subfloat[MPKI]{
       \includegraphics[width=0.95\columnwidth]{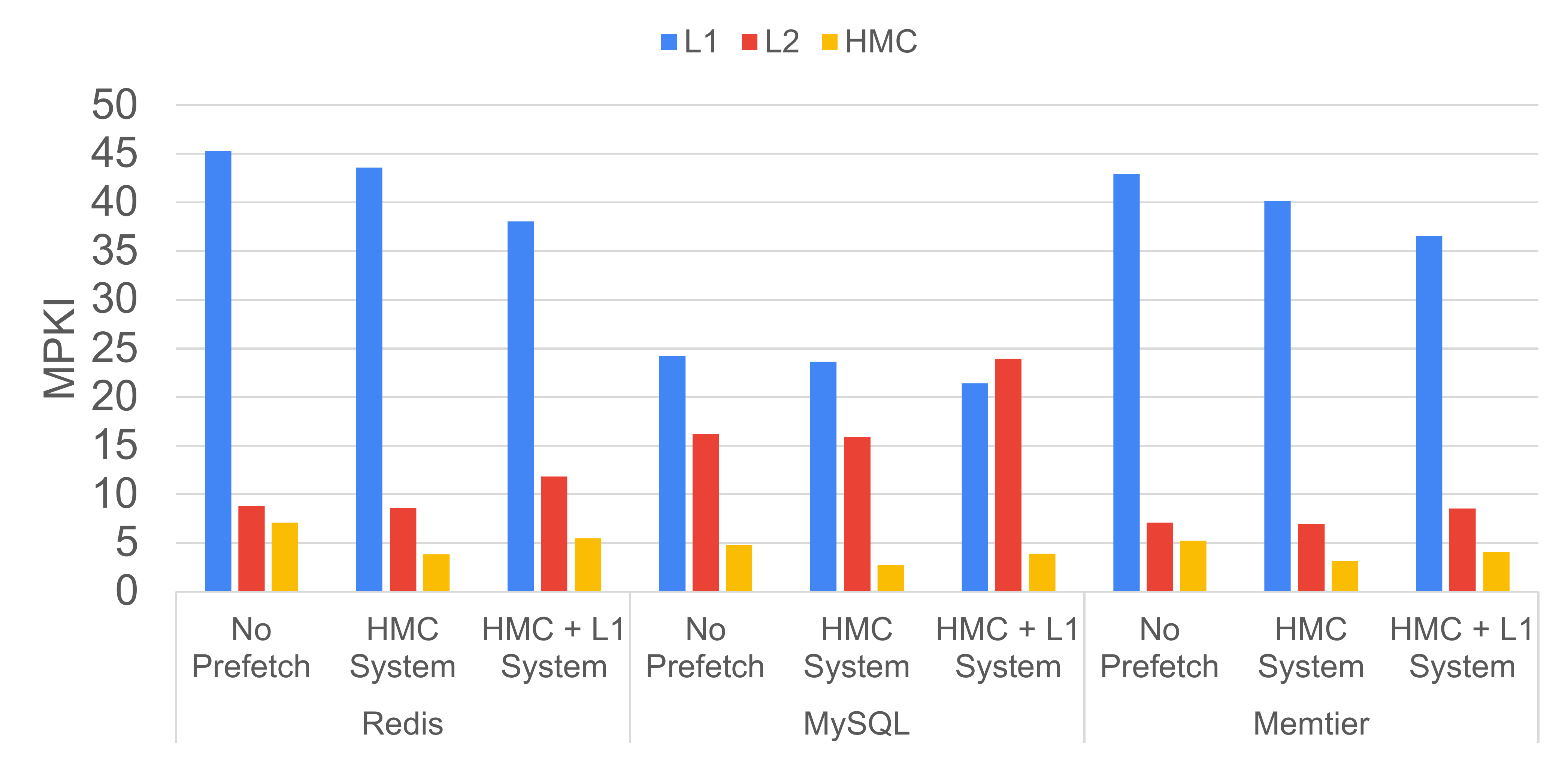}
       \label{fig:MPKI}
    }
    \subfloat[Miss latency]{
       \includegraphics[width=0.95\columnwidth]{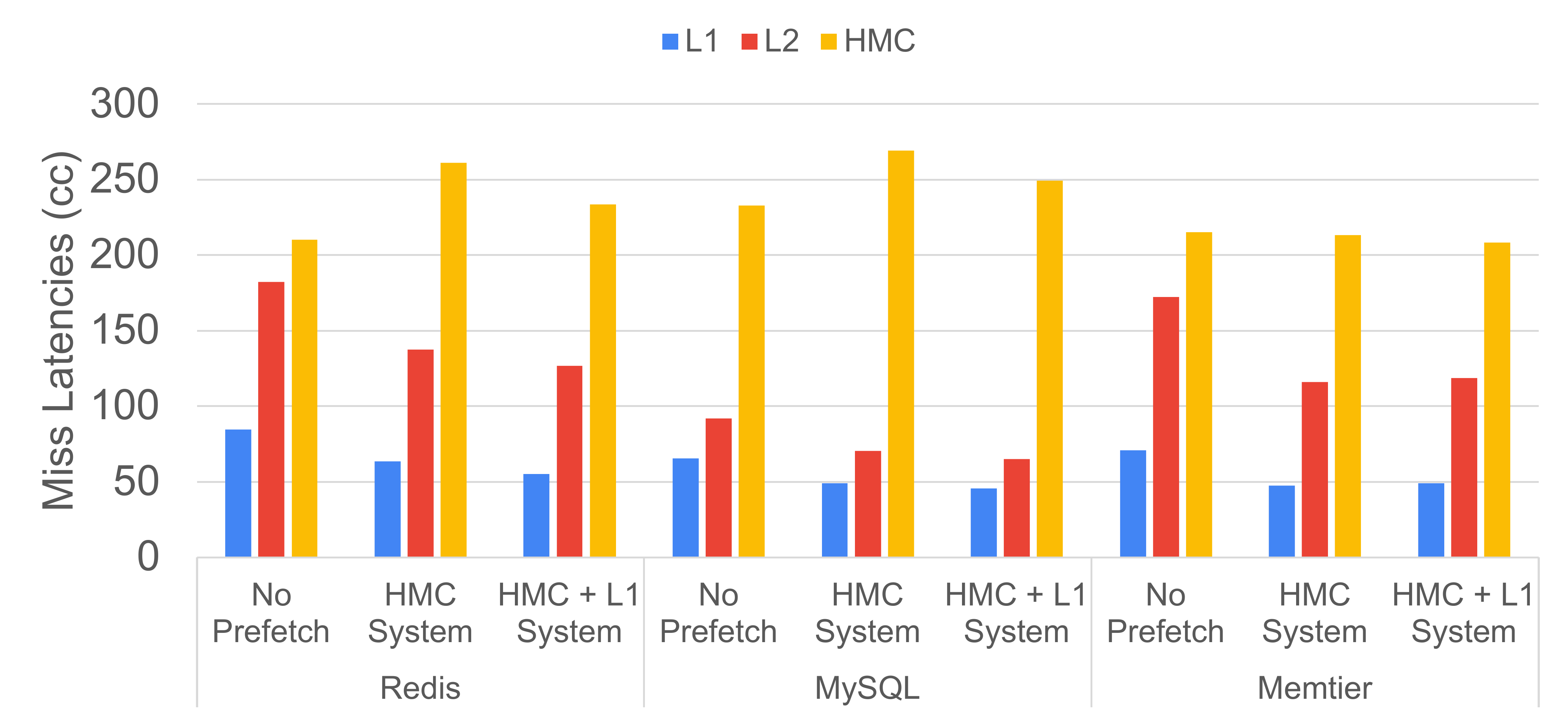}
       \label{fig:miss_latency}
    } 
    \\
    \centering
    \subfloat[AMAT]{
       \includegraphics[width=0.95\columnwidth]{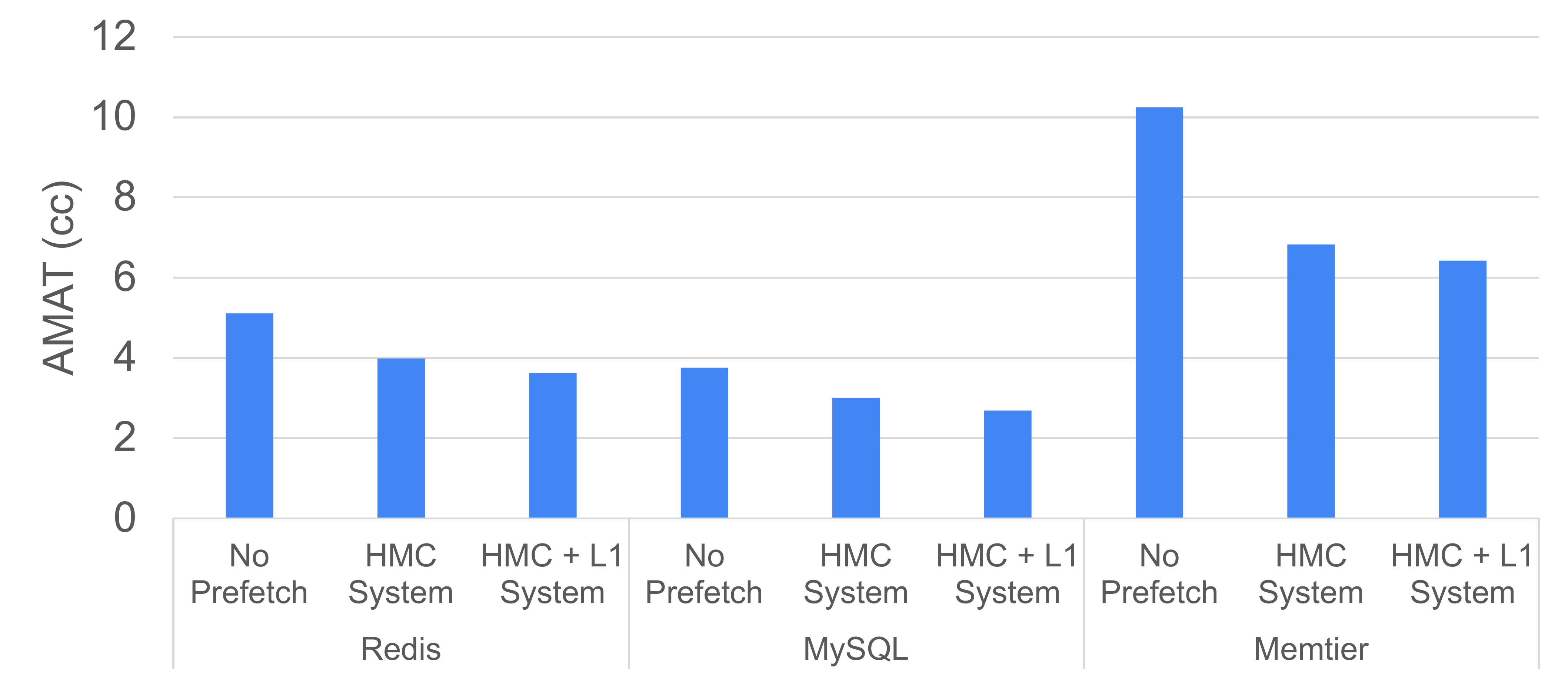}
       \label{fig:AMAT_o3}
    }
    \subfloat[IPC]{
       \includegraphics[width=0.95\columnwidth]{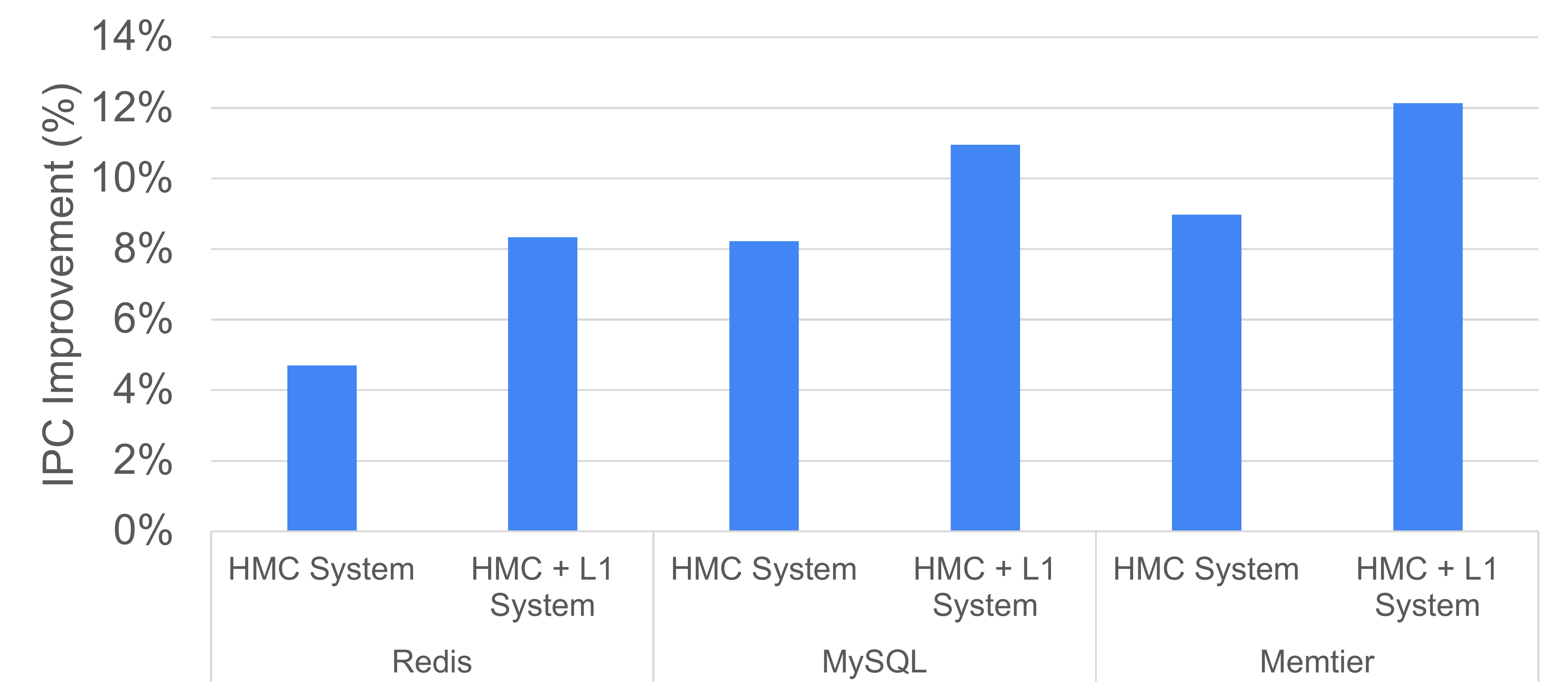}
       \label{fig:IPC_Imp_o3}
    }
    \caption{Memory hierarchy and overall performance metrics.}
    \label{fig:dynamic_predictions}
\end{figure*}

After checking that the prefeching systems present good coverage and accuracy, especially in the HMC cache,
we study to what extent these figures can improve the performance of the cache hierarchy. 
For this purpose, we analyze the impact of prefetching on the performance of each individual cache of the hierarchy (L1, L2, and HMC caches). This study concentrates on how prefetching affects  
the number of cache misses per kilo-instructions committed or MPKI and the cache miss penalty, which will be also indistinctively referred to as cache miss latency or merely cache latency. 

The study starts by analyzing the MPKI experienced by each cache organization of the hierarchy, which is compared to the MPKI experienced by a non-prefetching system also referred to as \emph{no prefetch}. We use the terms MPKI$_{L1}$, MPKI$_{L2}$, and MPKI$_{HMC}$, to refer the MPKI of the L1, L2, and HMC caches, respectively.
Figure \ref{fig:MPKI} shows the results for the studied systems.

Let us first analyze the impact on the HMC cache.
As observed, the MPKI of the HMC cache (MPKI$_{HMC}$) in the HMC prefetching system significantly improves over no prefetch (e.g., from about 6.8 to 4.2 in \texttt{redis}), confirming the excellent coverage results shown above. In the HMC+L1 system, however, the MPKI$_{HMC}$ is slightly higher.
The main reason is that this system also triggers prefetches at the top level of the cache hierarchy. More precisely, the (speculative) prefetches triggered by the L1 cache prefetches in the HMC+L1 system translate into a higher L2 cache miss ratio (and so higher MPKI$_{L2}$); therefore, more requests reach the HMC cache, which also translates into a higher HMC cache miss ratio.

An interesting observation is that the L1 cache prefetches (in the HMC+L1 system) play a key role as they significantly reduce MPKI$_{L1}$ (e.g. from 45 to 37 in \texttt{redis}). This is achieved thanks to anticipating L2 cache misses over both no prefetch and the HMC system.

In summary, the devised prefetching systems present good accuracy and coverage, which translates into significant MPKI improvements. Next, we analyze the cache miss latency.
Figure \ref{fig:miss_latency} presents the observed miss latency for the three SRAM caches (L1, L2, and HMC caches). As expected, the miss penalty of the HMC cache is higher in the prefetching systems
as more requests compete for access to the slow main NVRAM memory. 

It can be observed that both prefetching systems significantly improve the L2 cache miss latency over no prefetch, thanks to the coverage achieved in the HMC cache. In other words, many L2 cache misses result in a hit in the HMC cache, thus reducing the L2 miss latency.
For instance, L2 miss latency improves in the HMC system by 
33\% (from 182.15 to 137.33 processor cycles), 30\% 
and 48\% 
in \texttt{redis}, \texttt{MySQL}, and \texttt{memtier}, respectively. 
A similar rationale can be applied to L1 cache misses as it applies recursively.

\subsection{Overall Performance Analysis} 
\label{cap:overall-performance-analysis}


Once studied the impact of prefetching on the performance of the cache hierarchy, this section studies the impact on the overall system performance.

Among the analyzed latencies, in general, the one that shows the strongest relationship with the system performance is the L1 cache miss latency. This metric is the only one required to compute the AMAT (average memory access time), typically used to evaluate the performance of the memory hierarchy as a whole.
The AMAT can be computed with the following equation:
\[
AMAT = L1_{access\_time} + L1_{miss\_ratio} \times L1_{miss\_latency}
\]
Notice that $L1_{miss\_latency}$ can be in turn be computed as:
\[
L1_{miss\_latency} = L2_{access\_time} + L2_{miss\_ratio} \times L2_{miss\_latency}
\]

and $L2_{miss\_latency}$ can be also computed recursively. Notice, however, that the key metrics that need to be improved to increase the performance are the $L1_{miss\_latency}$ and the $L1_{miss\_ratio}$. The former metric summarizes how the memory hierarchy works up to reaching the bottom side of the hierarchy (i.e., the slowest NVRAM in our systems), whereas the latter determines the weight of the L1 miss latency in the AMAT.

Taking the previous rationale into account, 
%
%
%
we analyze the AMAT to provide a better understanding of how the performance of the memory subsystem impacts the overall system performance (i.e. instructions per cycle, IPC). 
Figure \ref{fig:AMAT_o3} presents the results for each benchmark across the studied systems. 
Results show that the HMC prefetching system improves AMAT over no prefetch by 28.50\% in \texttt{redis}, 24.99\% in \texttt{MySQL} and 50.08\% in \texttt{memtier}.
Including additional prefetchers in the L1 cache (HMC+L1 system) increases these figures up to 40.65\%, 39.97\%, and 59.36\%, respectively.
In the case of \texttt{redis} and \texttt{MySQL}, these improvements come from reductions in both L1 misses and the L1 miss latency, whereas \texttt{memtier} is mostly positively affected by a better L1 miss rate.

To what extent AMAT improvements translate into reduced execution time for a given application strongly depends on the o-o-o processor features that affect the \emph{computation to memory} ratio,
which is mainly established by the processor's aggressiveness. 
In other words, the higher the processor aggressiveness the higher the stress on the main memory as more memory accesses will be issued per unit time, and consequently, higher potential performance gains.

Thus, despite previous studies providing insights on performance improvements in different parts of the system, detailed simulation to reproduce the flow and rate at which memory instructions proceed is needed in order to reproduce memory level parallelism, and get the system performance (i.e. IPC).
In the modeled o-o-o processor, as depicted in Figure \ref{fig:IPC_Imp_o3}, performance gains in terms of speedup are as much as 4.71\%, 8.22\%, and 8.97\% in \texttt{redis}, \texttt{MySQL}, and \texttt{memtier} in the HMC system.

As observed, however, not all the workloads obtain a significant performance improvement just with the HMC system. This is the case of \texttt{redis}, where the achieved speedup is below 5\%.
The main reason is due to the off-chip latency (L2 latency in our experimental framework) is still relatively high (by 130 cycles, see Figure \ref{fig:miss_latency}), limiting the achievable performance (see Section \ref{sec:pref-systems-locations}).
Notice that the HMC+L1 system avoids this performance constraint by leveraging additional prefetchers in the L1 cache, improving AMAT over the HMC system, which translates into performance gains up to $8.33$\%, $10.96$\%, and $12.14$\% in \texttt{redis}, \texttt{MySQL}, and \texttt{memtier}, respectively.

\section{Related Work}
\label{sec:related}

Implementing NVRAM media as the main memory is seen as a likely replacement for the DRAM counterpart. However, some challenges need to be solved before NVRAM takes over. These challenges are mainly related to the long NVRAM access time, which would damage the system performance. 
An important piece of research has concentrated on the memory hierarchy organization to deal with 
this issue.

Some researchers have focused on utilizing NVRAM and DRAM to deploy a flat memory address space. 
This approach usually involves migrating data between NVRAM and DRAM to achieve shorter latencies, extend NVRAM lifespan, or reduce energy consumption.
For instance, in \cite{rowbuffermigration}, the migration policy considers potential future row-buffer misses in NVRAM, relocating them to DRAM to minimize energy consumption and access latency.
PDRAM \cite{pdram} focuses on enhancing NVRAM lifespan by monitoring the number of page accesses and migrating pages to DRAM when a threshold is reached.

Another approach to address the long NVRAM latencies
is using a hybrid memory address space. This approach  combines different memory technologies (e.g., DRAM and NVRAM) in the same design, acting their sum as the main memory address space. 
Since DRAM is faster, some approaches propose to devote a fraction of the DRAM memory to cache NVRAM pages. 
Compared to typical DRAM memory \cite{qureshi2009scalable},
this approach presents longer access latency.
A major drawback of this approach lies in the storage and management of the tags of the DRAM cache, which 
is challenging for huge DRAM caches. 
For instance, a 1GB DRAM cache with 64-byte blocks requires 96MB of tag storage.
This issue has been addressed in several works \cite{loh11,DRAMTags}.
On the other hand, Hardware/Software Cooperative Caching (HSCC) \cite{hwswcoop} seeks to streamline DRAM cache management by shifting it to the software layers of the architecture. HSCC organizes the flat address space of a hybrid memory system into a memory hierarchy, providing cache filtering policies to improve DRAM cache efficiency. Additionally, it optimizes cache management through software-based mechanisms. This work uses large cache blocks (i.e., cache sectors) and a relatively small DRAM cache to deal with this shortcoming. 

Recent proposals allow for re-configurable hybrid memory architectures capable of toggling between flat-like and cache-like hierarchies. One such proposal is Transformer \cite{transformer}
Transformer migrates most accessed pages to DRAM from DRAM,
and it can change 
to a memory architecture where DRAM acts as a cache of NVRAM. However, the main shortcoming of Transformer is
the costly requirements for OS monitoring of memory accesses for effective page handling. 
Hybrid$^2$ \cite{hybrid2} attempts to combine both caching and migration by establishing a DRAM cache from a fraction of the available DRAM media operating at sector granularity. On eviction, whether to migrate a page to DRAM media or not is decided. 
Finally, one notable commercial implementation of this togglable hierarchy was Intel's \emph{Optane Persistent Memory (PMem)}
\cite{Inteloptanepersistent}. This memory organization allows for specialized Intel Optane DIMMs to be configurable in three different modes: Memory Mode (DRAM as NVRAM cache), App Direct Mode (application-managed flat address space), and Mixed Mode (Memory and App Direct modes applied to different parts of the address space). 

Regarding prefetching, to the best of our knowledge, no work has been published about the impact of prefetching in a system combining on-chip and off-chip caches at the memory controller with different memory media.

\section{Conclusions}
\label{sec:conclusions}

Emergent applications require increasing amounts of main memory exceeding the capacity of current commodity processors equipped with DRAM technology as main memory. A possible solution to this issue is to move the DRAM main memory to the much denser NVRAM technology. This scenario involves, among others, research to concentrate on new memory controllers, additional off-chip caches, and prefetchers to hide the long NVRAM latencies.

This work has concentrated on analyzing the impact of prefetching on the performance of future computer systems working with multiple memory technologies. For this purpose, an off-chip memory controller that provides the processor access to distinct types of memory media (i.e., DRAM and NRAM) is used as the baseline system.
The main aim of this paper is to focus on a practical product.
Therefore, for evaluation purposes, we implemented prefetchers commonly adopted by processor manufacturers in commercial products. Two main systems have been devised: the HMC and the HMC+L1 prefetching system. The former implements a prefetcher at the HMC cache, and the latter extends the former system with additional L1 cache prefetchers.
Experimental results on an aggressive o-o-o processor show that the HMC prefeching system, by prefetching NVRAM sectors in the HMC cache presents coverage and accuracy results over
60\% in \texttt{redis} and 70\% in \texttt{memtier},
allowing overall performance improvement up to 9\%.
These results are further improved by the HMC+L1 prefetching system 
with coverages and performance improvement up to 92\% and 12\%, respectively.

\section*{Acknowledgment}

This work has been supported by the Spanish Ministerio de Ciencia e Innovación and European ERDF under grants PID2021-123627OB-C51, PID2024-158682OB-C31, by MCIN/AEI/10.13039/501100011033 and European Union NextGenerationEU/PRTR under grant TED2021-130233B-C32, and Generalitat Valenciana under grant CIPROM/2024/007. 

\bibliographystyle{IEEEtran}
\bibliography{IEEEabrv,DSD2025.bib}

\begin{thebibliography}{10}
\providecommand{\url}[1]{#1}
\csname url@samestyle\endcsname
\providecommand{\newblock}{\relax}
\providecommand{\bibinfo}[2]{#2}
\providecommand{\BIBentrySTDinterwordspacing}{\spaceskip=0pt\relax}
\providecommand{\BIBentryALTinterwordstretchfactor}{4}
\providecommand{\BIBentryALTinterwordspacing}{\spaceskip=\fontdimen2\font plus
\BIBentryALTinterwordstretchfactor\fontdimen3\font minus \fontdimen4\font\relax}
\providecommand{\BIBforeignlanguage}[2]{{%
\expandafter\ifx\csname l@#1\endcsname\relax
\typeout{** WARNING: IEEEtran.bst: No hyphenation pattern has been}%
\typeout{** loaded for the language `#1'. Using the pattern for}%
\typeout{** the default language instead.}%
\else
\language=\csname l@#1\endcsname
\fi
#2}}
\providecommand{\BIBdecl}{\relax}
\BIBdecl

\bibitem{Mak18}
H.~Mohammadi~Makrani, S.~Rafatirad, A.~Houmansadr, and H.~Homayoun, ``Main-memory requirements of big data applications on commodity server platform,'' in \emph{18th IEEE/ACM International Symposium on Cluster, Cloud and Grid Computing (CCGRID)}.\hskip 1em plus 0.5em minus 0.4em\relax IEEE, 2015, pp. 653--660.

\bibitem{Meza15}
M.~O., ``Revisiting memory errors in large-scale production data centers: Analysis and modeling of new trends from the field,'' in \emph{45th Annual IEEE/IFIP International Conference on Dependable Systems and Networks}.\hskip 1em plus 0.5em minus 0.4em\relax IEEE, 2015, pp. 415--426.

\bibitem{Mutlu16}
------, ``Rethinking memory system design,'' in \emph{Mobile System Technologies Workshop}, 2016, pp. 1--3.

\bibitem{IntelXeonPlatinum8470}
\BIBentryALTinterwordspacing
Intel. Intel® xeon® platinum 8470n processor. [Online]. Available: \url{https://www.intel.com/content/www/us/en/products/sku/231748/intel-xeon-platinum-8470n-processor-97-5m-cache-1-70-ghz/specifications.html}
\BIBentrySTDinterwordspacing

\bibitem{Burr08}
G.~Burr, B.~Kurdi, C.~Scott, C.~Lam, K.~Gopalakrishnan, and R.~Shenoy, ``Overview of candidate device technologies for storage-class memory,'' \emph{IBM Journal of Research and Development}, vol.~52, pp. 449--464, 07 2008.

\bibitem{Optane19}
A.~Altman, M.~Arafa, K.~Balasubramanian, K.~Cheng, P.~Damle, S.~Datta, C.~Douglas, K.~Gibson, B.~Graniello, J.~Grooms \emph{et~al.}, ``Intel optane data center persistent memory,'' \emph{IEEE Hot Chips 31 Symposium (HCS)}, pp. i--xxv, August 2019.

\bibitem{Optane21}
Y.~Yang, Q.~Cao, and S.~Wang, ``A comprehensive empirical study of file systems on optane persistent memory,'' \emph{IEEE International Conference on Networking, Architecture and Storage (NAS)}, pp. 1--8, 2021.

\bibitem{Hoya19}
K.~Hoya, K.~Hatsuda, K.~Tsuchida, Y.~Watanabe, Y.~Shirota, and T.~Kanai, ``A perspective on nvram technology for future computing system,'' in \emph{2019 International Symposium on VLSI Technology, Systems and Application (VLSI-TSA)}, 2019, pp. 1--2.

\bibitem{Inteloptanebusiness}
\BIBentryALTinterwordspacing
Intel. Intel® optane™ business update: What does this mean for warranty and support. [Online]. Available: \url{https://www.intel.com/content/www/us/en/support/articles/000091826/ memory-and-storage.html}
\BIBentrySTDinterwordspacing

\bibitem{Pel14}
S.~Pelley, P.~M. Chen, and T.~F. Wenisch, ``Memory persistency,'' in \emph{2014 ACM/IEEE 41st International Symposium on Computer Architecture (ISCA)}, 2014, pp. 265--276.

\bibitem{Zuo18}
P.~Zuo, Y.~Hua, M.~Zhao, W.~Zhou, and Y.~Guo, ``Improving the performance and endurance of encrypted non-volatile main memory through deduplicating writes,'' in \emph{2018 51st Annual IEEE/ACM International Symposium on Microarchitecture (MICRO)}.\hskip 1em plus 0.5em minus 0.4em\relax IEEE, 2018, pp. 442--454.

\bibitem{Luo18}
Y.~Luo, S.~Ghose, Y.~Cai, E.~F. Haratsch, and O.~Mutlu, ``Improving 3d {NAND} flash memory lifetime by tolerating early retention loss and process variation,'' \emph{Abstracts of the {ACM} International Conference on Measurement and Modeling of Computer Systems, {SIGMETRICS}, Irvine, CA, USA}, p. 106, June 2018.

\bibitem{Kagar22}
S.~Kargar and F.~Nawab, ``Challenges and future directions for energy, latency, and lifetime improvements in nvms,'' \emph{Distributed and Parallel Databases}, pp. 442--454, September 2022.

\bibitem{Avargues23}
M.~A. Avargues, M.~Lurbe, S.~Petit, M.~E. G{\'{o}}mez, R.~Yang, X.~Zhu, G.~Wang, and J.~Sahuquillo, ``Main memory controller with multiple media technologies for big data workloads,'' \emph{J. Big Data}, vol.~10, no.~1, p.~75, 2023.

\bibitem{rowbuffermigration}
H.~Yoon, J.~Meza, R.~Ausavarungnirun, R.~A. Harding, and O.~Mutlu, ``Row buffer locality aware caching policies for hybrid memories,'' in \emph{2012 IEEE 30th International Conference on Computer Design (ICCD)}.\hskip 1em plus 0.5em minus 0.4em\relax IEEE, 2012, pp. 337--344.

\bibitem{pdram}
G.~Dhiman, R.~Ayoub, and T.~Rosing, ``Pdram: A hybrid pram and dram main memory system,'' in \emph{2009 46th ACM/IEEE Design Automation Conference}, 2009, pp. 664--669.

\bibitem{qureshi2009scalable}
M.~K. Qureshi, V.~Srinivasan, and J.~A. Rivers, ``Scalable high performance main memory system using phase-change memory technology,'' in \emph{Proceedings of the 36th annual international symposium on Computer architecture}, 2009, pp. 24--33.

\bibitem{loh11}
G.~H. Loh and M.~D. Hill, ``Efficiently enabling conventional block sizes for very large die-stacked dram caches,'' in \emph{Proceedings of the 44th Annual IEEE/ACM International Symposium on Microarchitecture}, ser. MICRO-44.\hskip 1em plus 0.5em minus 0.4em\relax New York, NY, USA: Association for Computing Machinery, 2011, p. 454–464.

\bibitem{DRAMTags}
M.~K. Qureshi and G.~H. Loh, ``Fundamental latency trade-off in architecting dram caches: Outperforming impractical sram-tags with a simple and practical design,'' in \emph{2012 45th Annual IEEE/ACM International Symposium on Microarchitecture}, 2012, pp. 235--246.

\bibitem{transformer}
Y.~Chi, H.~Liu, G.~Peng, X.~Liao, and H.~Jin, ``Transformer: An os-supported reconfigurable hybrid memory architecture,'' \emph{Applied Sciences}, vol.~12, no.~24, 2022.

\bibitem{hybrid2}
E.~Vasilakis, V.~Papaefstathiou, P.~Trancoso, and I.~Sourdis, ``Hybrid2: Combining caching and migration in hybrid memory systems,'' in \emph{2020 IEEE International Symposium on High Performance Computer Architecture (HPCA)}.\hskip 1em plus 0.5em minus 0.4em\relax IEEE, 2020, pp. 649--662.

\bibitem{bhatia19}
E.~Bhatia, G.~Chacon, S.~Pugsley, E.~Teran, P.~V. Gratz, and D.~A. Jim\'{e}nez, ``Perceptron-based prefetch filtering,'' in \emph{Proceedings of the 46th International Symposium on Computer Architecture}, ser. ISCA '19.\hskip 1em plus 0.5em minus 0.4em\relax New York, NY, USA: Association for Computing Machinery, 2019, p. 1–13.

\bibitem{bakhshalipour19}
M.~Bakhshalipour, M.~Shakerinava, P.~Lotfi-Kamran, and H.~Sarbazi-Azad, ``Bingo spatial data prefetcher,'' in \emph{2019 IEEE International Symposium on High Performance Computer Architecture (HPCA)}, 2019, pp. 399--411.

\bibitem{shevgoor15}
M.~Shevgoor, S.~Koladiya, R.~Balasubramonian, C.~Wilkerson, S.~H. Pugsley, and Z.~Chishti, ``Efficiently prefetching complex address patterns,'' in \emph{2015 48th Annual IEEE/ACM International Symposium on Microarchitecture (MICRO)}, 2015, pp. 141--152.

\bibitem{pakalapati20}
S.~Pakalapati and B.~Panda, ``Bouquet of instruction pointers: Instruction pointer classifier-based spatial hardware prefetching,'' in \emph{2020 ACM/IEEE 47th Annual International Symposium on Computer Architecture (ISCA)}, 2020, pp. 118--131.

\bibitem{michaud16}
P.~Michaud, ``Best-offset hardware prefetching,'' in \emph{2016 IEEE International Symposium on High Performance Computer Architecture (HPCA)}, 2016, pp. 469--480.

\bibitem{kim16}
J.~Kim, S.~H. Pugsley, P.~V. Gratz, A.~N. Reddy, C.~Wilkerson, and Z.~Chishti, ``Path confidence based lookahead prefetching,'' in \emph{2016 49th Annual IEEE/ACM International Symposium on Microarchitecture (MICRO)}, 2016, pp. 1--12.

\bibitem{navarro22}
A.~Navarro-Torres, B.~Panda, J.~Alastruey-Benedé, P.~Ibáñez, V.~Viñals-Yúfera, and A.~Ros, ``Berti: an accurate local-delta data prefetcher,'' in \emph{2022 55th IEEE/ACM International Symposium on Microarchitecture (MICRO)}, 2022, pp. 975--991.

\bibitem{arm-L1Icache-next-line}
``{arm Developer} hardware l1 i-cache prefetching,'' \url{https://developer.arm.com/documentation/ddi0488/h/level-1-memory-system/l1-instruction-memory-system/hardware-l1-i-cache-prefetching}, accessed: 2022-12-13.

\bibitem{arm-L2-next-line}
``{arm Developer} l2 cache prefetcher,'' \url{https://developer.arm.com/documentation/ddi0488/h/level-2-memory-system/l2-cache-prefetcher}, accessed: 2022-12-13.

\bibitem{corporation2018intel}
Intel, ``Intel 64 and ia-32 architectures opt. reference manual,'' 2018.

\bibitem{tabla-stride}
T.-F. Chen and J.-L. Baer, ``Effective hardware-based data prefetching for high-performance processors,'' \emph{IEEE Transactions on Computers}, vol.~44, no.~5, pp. 609--623, 1995.

\bibitem{corporation2022intel}
Intel, ``Intel 64 and ia-32 architectures opt. reference manual,'' 2022.

\bibitem{bajaber20}
\BIBentryALTinterwordspacing
F.~Bajaber, S.~Sakr, O.~Batarfi, A.~Altalhi, and A.~Barnawi, ``Benchmarking big data systems: A survey,'' \emph{Computer Communications}, vol. 149, pp. 241--251, 2020. [Online]. Available: \url{https://www.sciencedirect.com/science/article/pii/S0140366419312344}
\BIBentrySTDinterwordspacing

\bibitem{panda17}
R.~Panda and L.~K. John, ``Proxy benchmarks for emerging big-data workloads,'' in \emph{2017 26th International Conference on Parallel Architectures and Compilation Techniques (PACT)}, 2017, pp. 105--116.

\bibitem{dawodi19}
M.~Dawodi, M.~H. Hedayati, J.~A. Baktash, and A.~L. Erfan, ``Facebook mysql performance vs mysql performance,'' in \emph{2019 IEEE 10th Annual Information Technology, Electronics and Mobile Communication Conference (IEMCON)}, 2019, pp. 0103--0109.

\bibitem{cacti7.0}
N.~Muralimanohar, R.~Balasubramonian, and N.~P. Jouppi, ``Cacti 6.0: A tool to model large caches,'' \emph{HP laboratories}, vol.~27, p.~28, 2009.

\bibitem{hwswcoop}
H.~Liu, Y.~Chen, X.~Liao, H.~Jin, B.~He, L.~Zheng, and R.~Guo, ``Hardware/software cooperative caching for hybrid dram/nvm memory architectures,'' in \emph{Proceedings of the International Conference on Supercomputing}, 2017, pp. 1--10.

\bibitem{Inteloptanepersistent}
\BIBentryALTinterwordspacing
Intel. Why is the intel® optane™ persistent memory in memory mode not persistent? [Online]. Available: \url{https://www.intel.com/content/www/us/en/support/articles/000055895/ memory-and-storage/intel-optane-persistent-memory.html}
\BIBentrySTDinterwordspacing

\end{thebibliography}

\end{document}